\documentclass{article}
\usepackage[utf8]{inputenc}
\usepackage[a4paper,top=2cm,bottom=2cm,left=3cm,right=3cm,marginparwidth=1.75cm]{geometry}
\usepackage{gensymb}
\usepackage{graphicx}
\usepackage[colorlinks=true, allcolors=blue]{hyperref}
\usepackage{authblk}
\usepackage{amsmath}
\usepackage[english]{babel}
\usepackage{titlesec}
\usepackage{xcolor}
\usepackage{pdfpages}
\titleformat{\paragraph}
{\normalfont\normalsize\bfseries}{\theparagraph}{1em}{}
\titlespacing*{\paragraph}
{0pt}{3.25ex plus 1ex minus .2ex}{1.5ex plus .2ex}
\usepackage[
    backend=biber,
    style=nature,
  ]{biblatex}

\title{Signature of lattice dynamics in twisted 2D homo/hetero-bilayers}

\author[1,2]{Yang Pan}
\author[3]{Shutong Li}
\author[1,2*]{Mahfujur Rahaman}
\author[1,2]{Ilya Milekhin}
\author[1,2*]{\\Dietrich R. T. Zahn}
\affil[1]{Semiconductor Physics, Institut für Physik, Chemnitz University of Technology, Chemnitz, Germany}
\affil[2]{Center for Materials, Architectures, and Integration of Nanomembranes (MAIN), Chemnitz University of Technology, Chemnitz, Germany}
\affil[3]{Department of Chemical Engineering and Materials Science, University of Minnesota, Minneapolis, Minnesota, USA}
\affil[*]{Corresponding author: Dietrich R. T. Zahn: zahn@physik.tu-chemnitz.de

Mahfujur Rahaman}

\date{}

\begin{document}

\maketitle

\begin{abstract}

Twisted 2D bilayer materials are created by artificial stacking of two monolayer crystal networks of 2D materials with a desired twisting angle $\theta$. The material forms a moiré superlattice due to the periodicity of both top and bottom layer crystal structure. The optical properties are modified by lattice reconstruction and phonon renormalization, which makes optical spectroscopy an ideal characterization tool to study novel physics phenomena. Here, we report a Raman investigation on a full period of the twisted bilayer (tB) WSe$_2$ moiré superlattice (\textit{i.e.} 0° $\leq \theta \leq$ 60°). We observe that the intensity ratio of two Raman peaks, $B_{2g}$ and $E_{2g}/A_{1g}$ correlates with the evolution of moiré period. The Raman intensity ratio as a function of twisting angle follows an exponential profile matching the moiré period with two local maxima at 0° and 60° and a minimum at 30°. Using a series of temperature-dependent Raman and photoluminescence (PL) measurements as well as \textit{ab initio} calculations, the intensity ratio  $I_{B_{2g}}/I_{{E_{2g}}/{A_{1g}}}$ is explained as a signature of lattice dynamics in tB WSe$_2$ moiré superlattices. By further exploring different material combinations of twisted hetero-bilayers, the results are extended for all kinds of Mo- and W-based TMDCs.

\end{abstract}

\section{Introduction}

Due to the weak van der Waals interlayer coupling, 2D materials enable the possibility to stack one layer onto the other with arbitrary, yet precisely controlled relative crystal orientations \cite{geim2013van}. Vertical stacking of homo- or hetero-bilayer 2D materials leads to the formation of moiré superlattices, which induce periodic modulations, potential distribution, phonon renormalization, lattice reconstruction, and optical selection rules  \cite{tran2019evidence,zhang2017interlayer,li2021imaging,quan2021phonon,gadelha2021localization,yu2021imaging}. The reconstructed moiré superlattice provides an exciting platform to investigate novel physics phenomena. The moiré system in 2D materials promises a wide range of potential applications in electronic and optoelectronic devices \cite{koppens2014photodetectors,xie2017photodetectors,yin2016selectively,cao2016gate}. For example, the moiré superlattice generates flat minibands with a high density of states at the edge of the original Brillouin zones (BZ) of the stacked bilayer systems (also known as moiré BZ) \cite{ruiz2019interlayer}. Cao \textit{et al}. reported superconductivity and correlated insulator behavior in magic-angle twisted bilayer graphene \cite{cao2018unconventional,cao2018correlated}, which attracted tremendous research attention to other twisted 2D material systems. Especially, interlayer excitons in TMDC heterobilayers can be affected by the moiré potential due to the fluctuation of the atomic registries in the moiré supercell leading to the formation of 2D arrays of identical quantum wells on the nanoscale \cite{seyler2019signatures,brem2020tunable}.

Twisted 2D systems have a promising future, yet the lattice dynamics is not thoroughly understood, and it is still challenging to characterize the twisting angle effectively and precisely. Researchers have developed characterization methods based on different techniques. On the one hand, characterization methods such as transmission electron microscopy (TEM) \cite{latychevskaia2019moire}, scanning tunneling microscopy (STM) \cite{li2021imaging,flores2013moire}, piezoresponse force microscopy (PFM) \cite{mcgilly2020visualization}, Kelvin probe force microscopy (KPFM) \cite{yu2021imaging}, microwave impedance microscopy (MIM) \cite{huang2021imaging}, conductive atomic force microscopy (CAFM) \cite{rosenberger2020twist}, and tip-enhanced Raman spectroscopy (TERS) \cite{gadelha2021localization} can resolve the moiré superlattice on the nanoscale, however, all share the drawback of the complexity of the respective experiments often accompanied by high time consumption. On the other hand, second harmonic generation (SHG) provides rich information in 2D materials and has been widely used to characterize both homo- and hetero-structured 2D materials. However, SHG is limited due to the relative complicated optical setup and lack of information on phonons \cite{ma2020rich}.  Raman spectroscopy is a reliable and non-destructive method to investigate lattice vibrations/phonon modes of materials. Low-frequency Raman spectra of twisted 2D systems have been intensively studied, since the interlayer shearing and layer breathing modes are sensitive to the coupling of layered materials and hence provide a means of characterization of such systems \cite{quan2021phonon,saito2016raman,huang2016low,parzefall2021moire}. However, the low frequency Raman modes have the drawback of relatively weak intensity and disappearing of modes at certain twisting angle \cite{quan2021phonon,parzefall2021moire}. Moreover, 2D systems are usually prepared on hBN substrates and the low frequency parts also include oscillatory signals originating from the coupling between the twisted 2D systems and the hBN substrate \cite{quan2021phonon,lin2019cross}, thus making data processing extremely challenging.

The high frequency Raman features in twisted 2D systems, on the other hand, are not yet studied thoroughly. Here, we present a Raman spectroscopic study on a full period of twisted bilayer (tB) WSe$_2$ (twisting angle 0° $\leq \theta \leq$ 60°) as a prototype of twisted transition metal dichalcogenide (TMDC) systems. Interestingly, we observe that the intensity ratio $I_{B_{2g}}/I_{{E_{2g}}/{A_{1g}}}$ of the $B_{2g}$ and $E_{2g}/A_{1g}$ Raman modes correlates well with the evolution of moiré period. The intensity ratio has two asymmetric local maxima at 0° and 60° twisting angle, and in between it follows exponential profiles matching the same trend as the progression of moiré superlattice constant with a local minimum at 30° twisting angle. Most interestingly, the $B_{2g}$ mode disappears at 30° twisting angle, \textit{i.e.} the Raman spectrum of tB WSe$_2$ then resembles that of monolayer WSe$_2$. From a series of temperature-dependent Raman measurements, we observe a completely different evolution of the spectra for intrinsic bilayer (iB) WSe$_2$ and 30° tB WSe$_2$. The Raman signal of iB WSe$_2$ almost vanishes when the temperature is higher than 473 K, while the Raman signal for 30° tB and monolayer WSe$_2$ remains visible. This phenomenon is explained by the resonant Raman condition in good agreement with Ref. \cite{zhao2013evolution,del2014excited}. We also performed \textit{ab initio} calculations of the lattice dynamics of tB WSe$_2$ (0° $\leq \theta \leq$ 60°) to understand the effect of twisting angle dependent interlayer coupling on the first order Raman modes of WSe$_2$. Our work demonstrates that the intensity ratio $I_{B_{2g}}/I_{{E_{2g}}/{A_{1g}}}$ can be used as a signature of lattice dynamics in the tB system. We also explore different combination of twisted hetero-bilayer (thB) TMDC systems to corroborate our hypothesis. In general, our work demonstrates the great potential of high frequency micro-Raman spectroscopy to provide information on highly localized interlayer coupling, phonon renormalization, and lattice reconstruction.

\section{Materials and methods}

\subsection{Sample preparation}

Few layers of hBN are mechanically exfoliated from bulk material (epitaxial solidification technique grown, purchased from 2D semiconductors) and transferred onto a Si substrate with 300 nm SiO$_2$ via Nitto tape. Monolayer WSe$_2$ is exfoliated from bulk 2H phase WSe$_2$ (from HQ Graphene) onto polydimethylsiloxane (PDMS) by Nitto tape followed by a deterministic all-dry transfer technique \cite{castellanos2014deterministic}. A tear-and-stack method \cite{kim2016van} is performed to fabricate tB WSe$_2$. A large area of a monolayer WSe$_2$ flake is first exfoliated onto PDMS. Bringing the monolayer WSe$_2$ flake partially in contact with hBN and then slowly peeling off leaves the monolayer WSe$_2$ partially on hBN while the rest part still sticks on the PDMS stamp. In the next step a homo-bilayer sample is prepared by rotating the substrate with a defined angle and bringing the remaining part of the WSe$_2$ flake to contact. Via this procedure, the twisting angle can be well defined. The procedure is shown in detail in Fig. S1.

Twisted hetero-bilayers are prepared by an advanced tear-and-stack method. After preparing a 30° twisted homo-bilayer, the top layer of different TMDC materials is stamped via PDMS. The edge of the top layer is aligned to any of the edges of the bottom homo-bilayer with an angle of 30\textdegree $\times n$ ($n$ is any integer number since monolayer TMDCs belong to the D$_{3h}$ point group). This results in two areas of heterostructures, one with a relative twisting angle of 30° and the other with 0° or 60°. The procedure is shown in detail in Fig. S2.

\subsection{Optical spectroscopy}

Temperature-dependent Raman and PL measurements are conducted in a Linkam THMSEL600 chamber under a 50x, 0.5 NA objective, while room temperature Raman spectra are measured in ambient condition under a 100x, 0.9 NA objective. Temperature-dependent PL and room temperature Raman measurements are performed using a Horiba Xplora Plus equipped with a spectrometer comprising 600 l/mm (for PL) and 2400 l/mm (for Raman, with 1.6 cm$^{-1}$ spectral resolution) gratings  and an electron multiplying CCD (EMCCD). A DPSS 532 nm CW laser source was used to excite the samples with an excitation power of 100 µW measured under the objective. Temperature-dependent Raman spectra are acquired by a Horiba LabRAM HR spectrometer with a 2400 l/mm grating and an excitation wavelength of 514.7 nm with a spectral resolution of 0.8 cm$^{-1}$.

\subsection{First-principles calculation}

Density functional theory calculations are performed using the projector augmented wave approach as implemented in the Vienna \textit{ab initio} Simulation Package ({\sc vasp}) \cite{Blochl1994, Kresse1996, Kresse1999}. The DFT-D3 method \cite{Grimme2010} is employed to capture the interlayer Van der Waals forces accurately, which is crucial in the bilayer system. PBEsol \cite{PBEsol} is used as the approximation functional for the exchange-correlation functional and the cutoff energy is set to be $450~eV$. For 0° and 60° twisted systems, the unit cells consist of two triangular prisms. A 20$\AA$-thick vacuum layer is set in the unit cell to simulate the boundary conditions. A k-grid of 16 x 16 x 1 points is used for these two configurations. A set of commensurate twisted bilayer WSe$_2$ structures are also calculated with various numbers of atoms, including those with twisting angles of 21.8°, 27.8°, and 38.2° \cite{huang2016low}. A smaller k-grid of 4 x 4 x 1 points is used for these configurations because of the much smaller reciprocal lattice. All structures are relaxed until all forces on individual atoms are below $0.1~meV/\AA$. The phonon modes and their frequencies are determined using finite difference method. Density functional perturbation theory is also used to determine the ion-clamped dielectric constant \cite{DFPT1, DFPT2}, which is used to determine the Raman tensor of selected phonon modes.

The Raman tensor $\Tilde{R}$ of a given phonon mode can be calculated by the change of polarizability \cite{phonopy-raman}:

\begin{equation}
   \Tilde{R}_{\alpha \beta} \propto \frac{\Delta \varepsilon^{\infty}_{\alpha\beta}}{\Delta Q(s)}
\end{equation}

where $\varepsilon$ stands for the ion-clamped dielectric constant, \textit{i.e.} the dielectric constant at infinite high frequency, $Q(s)$ stands for the amplitude of the normal mode $s$, $\alpha$ and $\beta$ indicate the directions. The derivative can be calculated by freezing in the phonon mode $s$ and calculate the change of the dielectric constant. The nonresonant Raman intensity is usually calculated by the Placzek approximation \cite{Placzek}: $I_{Raman}\propto |\vec{e_{i}}\cdot\Tilde{R}\cdot\vec{e_{s}}|$. The $\vec{e}_i$ and $\vec{e}_s$ are the polarization directions of incident and scattered light. In the experimental setup, the incident light is polarized in-plane, noted as $x$. Both directions of in-plane scattered light were collected, which results in a total Raman intensity $I\propto I_{xx} + I_{xy} \propto R_{xx}^2 + R_{xy}^2$.

\section{Results and discussion}

\subsection{Theoretical analysis}

In bulk WSe$_2$, there are four Raman-active modes according to group theory analysis (frequency from low to high \cite{luo2013effects}): $E_{2g}^2$, $E_{1g}$, $E_{2g}^1$ and $A_{1g}$. Among these phonon modes, the inter-layer shearing mode $E_{2g}^2$ has a low frequency of $35cm^{-1}$, because it is dominated by van der Waals interaction \cite{zhao2013lattice}. The intensity of $E_{1g}$ is low and hard to measure \cite{luo2013effects}. Thus, in the following sections, we mainly focus on the last two modes and another interesting mode $B_{2g}$, which becomes Raman active with the reduction of layers. The schematics of these three modes can be found in the insets of Fig.~\ref{fig:1_tB_Raman}(a): The $E_{2g}$ (short for $E_{2g}^1$) mode is an in-plane vibrational mode, where the W and Se atoms vibrate against each other. The $A_{1g}$ mode is known to be an out-of-plane vibrational mode, where the 2 Se atoms on the opposite sides of a W atom vibrate against each other while the W atom shows no relative motion. The $B_{2g}$ mode is also an out-of-plane one, for which the Se and W atoms within the same layer vibrate against each other with a 180° phase difference with regard to the vibration in the adjacent layer.

Bulk WSe$_2$ can be mechanically exfoliated into 60° twisted bilayer structure \cite{zhao2013lattice}, which is known as the intrinsic bilayer structure. During this process, the number of symmetry elements are reduced from 24 to 12, among which there is a mirror plane in the W atoms layer perpendicular to the c-axis \cite{luo2013effects}. The mirror plane is important because the $B_{2g}$ mode elongation is reversed when this c-axis mirror plane is applied but remains unchanged when the 3-fold rotation is applied, neither symmetry operation transforms it like the basis function of $yz$ or $xz$, which makes it Raman inactive. The reduction in symmetry elements makes the $B_{2g}$ mode Raman active. However, if the number of layers is reduced to one, the c-axis mirror plane symmetry makes the $B_{2g}$ mode inactive in the monolayer. It is noteworthy that the irreducible representations (irreps) for the same phonon mode might change in different structures, but we usually refer to these phonon modes by their irreps in the bulk structures to be consistent. A complete list of irreps of the phonon modes in different structures can be found in Table.~\ref{tab:irreps}.

When the bilayer structure is twisted, our group theory analysis shows the $B_{2g}$ mode remains active. At most twisting angles besides 0° and 60°, there are four symmetry elements: a 3-fold axis and three 2-fold axis perpendicular to each other. This 3-fold axis can be found where two tungsten atoms from different layers overlap in the z-direction. This overlapping certainly happens in a moiré pattern. The $B_{2g}$ mode remains unchanged under these symmetry operations, which makes it behave exactly the same way as the basis functions of $x^2+y^2$ and $z^2$. At 0°, this statement holds although multiple other symmetry elements are present.

In Table.~\ref{tab:irreps}, we list the phonon modes by their irreps in different structures, along with their Raman-activity and frequencies calculated from density functional theory. These simulated frequencies match fairly well with our experimental values in the next section, which confirms our mode assignment. Our phonon calculation also shows that, at an arbitrary twisting angle, the phonon modes that are originally not at the $\Gamma$-point can fold back and couple with the ones at the $\Gamma$-point. This phenomenon is significant for the in-plane $E_{2g}$ mode: it can split into a bunch of peaks that are close to each other, however, this is very hard to be observed experimentally.

\begin{table}[h!]
\centering
\begin{tabular}{|l|c|c|c|}
\hline
& Out-of-plane shearing & Out-of-plane breathing & In-plane shearing\\ \hline
Bulk, 2H (\#194) & $B_{2g}$: \textbf{inactive} & $A_{1g}$ & $E_{2g}$ \\ \hline
0°-tB, 3R (\#187) & $A_1$ & $A_1$: & $E$ \\
& 311.2  & 252.3 & 248.2 \\ \hline
21.8°-tB (\#150) & $A_1$ & $A_1$ & $E$ \\
& 311.4  & 251.6 & 248.2 - 249.3 \\ \hline
27.8°-tB (\#149) & $A_1$ & $A_1$ & $E$ \\ 
& 311.4  & 251.8 & 248.8 - 249.8 \\
38.2°-tB (\#149) & $A_1$ & $A_1$ & $E$ \\ 
& 311.4  & 251.5 & 248.1 - 249.2 \\ \hline
60°-tB, 2H (\#164) & $A_{1g}$ & $A_{1g}$ & $E_g$ \\
& 310.8  & 251.1 & 248.6 \\ \hline
Monolayer (\#156)& $A_2''$: \textbf{inactive} & $E'$ & $A_1'$\\
\hline
\end{tabular}
\caption{Phonon modes and their irreps, frequencies (in $cm^{-1}$) in different structures. Most of the phonon modes in this table are Raman-active, with a few exceptions indicated by the label ``inactive". The space group numbers are noted in the parentheses next to each structure name \cite{dresselhaus2007group,stokes2005findsym}. The frequencies of some models are given as a range. Details of these structures can be found in the supplementary materials.}
\label{tab:irreps}
\end{table}

\subsection{Twisted homo-bilayers}

\begin{figure} 
\makebox[\textwidth][c]{\includegraphics[width=1\textwidth]{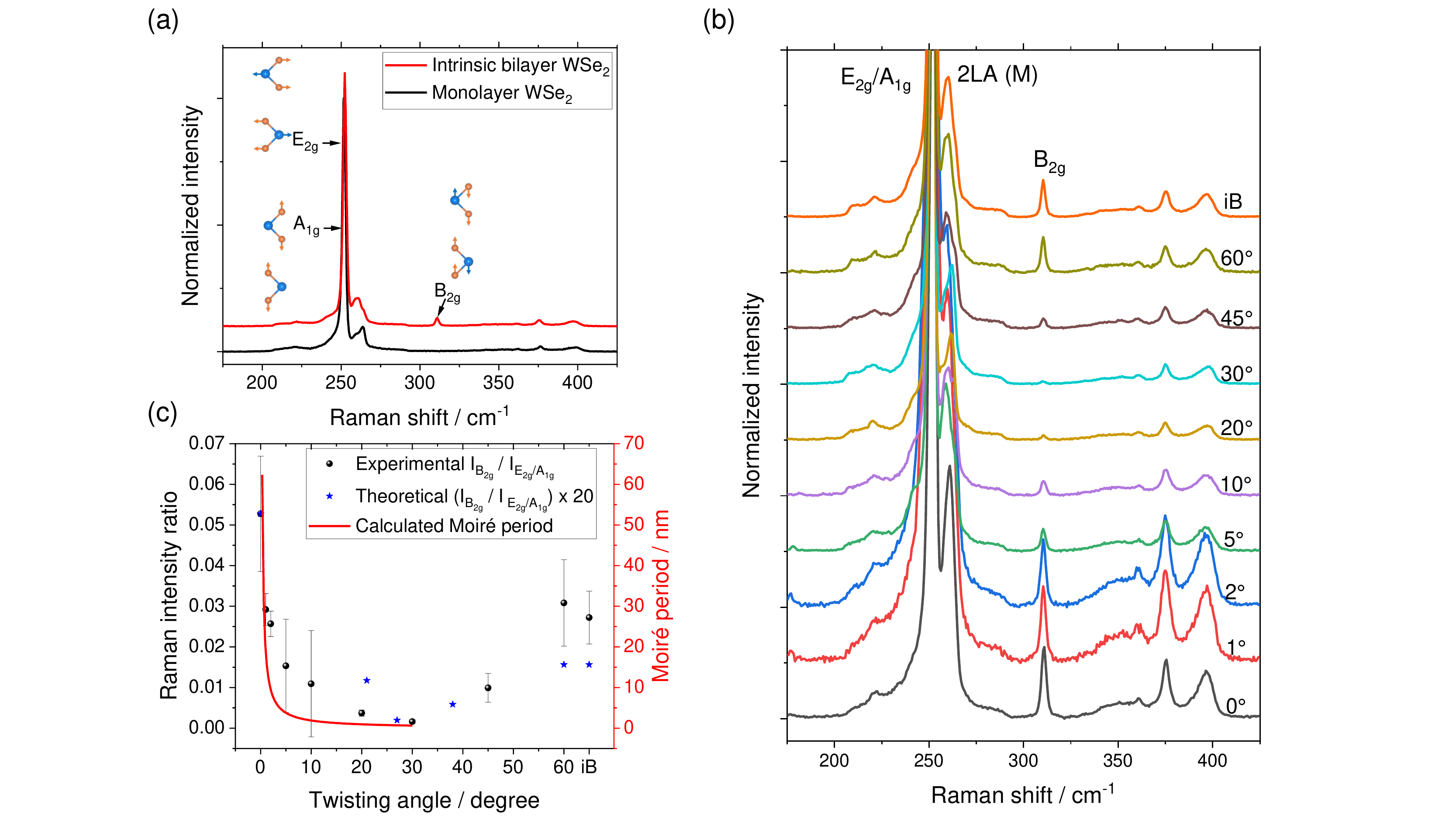}}  
\caption{(a) Raman spectra of monolayer and intrinsic bilayer WSe$_2$, the elongation patterns of the relevant observed modes are included. (b) Raman spectra of twisted bilayer WSe$_2$. All spectra are normalized with respect to the $E_{2g}/A_{1g}$ peak. (c) Calculated intensity ratio and experimental intensity ratio of $I_{B_{2g}}/I_{{E_{2g}}/{A_{1g}}}$ from (b), and calculated moiré superlattice constant.}  
\label{fig:1_tB_Raman}
\end{figure}

Fig. \ref{fig:1_tB_Raman}(a) shows the Raman spectra of monolayer and intrinsic bilayer WSe$_2$. The most pronounced peak corresponds to the combination of two vibrational modes, $E_{2g}$ and $A_{1g}$ mode, almost degenerate at around 250 $cm^{-1}$ \cite{luo2013effects,zhao2013lattice,tonndorf2013photoluminescence,terrones2014new}. The feature at around 260 $cm^{-1}$ is a second order peak caused by a double resonance effect involving the longitudinal acoustic phonon at the M point in the Brillouin zone assigned as 2LA (M) \cite{del2014excited,terrones2014new}. The most interesting feature is the out-of-plane $B_{2g}$ mode located at around 309 $cm^{-1}$. According to the group theory analysis, the $B_{2g}$ mode is Raman inactive in both bulk and monolayer, but becomes active in few layers due to the reduction of symmetry elements \cite{tonndorf2013photoluminescence,terrones2014new,ding2011first}.

Fig. \ref{fig:1_tB_Raman}(b) displays Raman spectra for a full period ($0\degree\leq\theta\leq60\degree$) tB WSe$_2$. Raman intensity fluctuations during different Raman measurements are eliminated by normalizing all spectra with respect to the most intense $E_{2g}/A_{1g}$ peak. With increasing twisting angle, $I_{B_{2g}}$ decreases and almost vanishes at 30°, then increases again and reaches a second maximum at 60°. All spectra are fitted with Voigt functions and the integrated areas are taken as a measure for the intensity. To visualize the effect the intensity ratio $I_{B_{2g}}/I_{{E_{2g}}/{A_{1g}}}$ is plotted as a function of twisting angle in Fig. \ref{fig:1_tB_Raman}(c). The red solid line in Fig. \ref{fig:1_tB_Raman}(c) is the calculated moiré superlattice constant (moiré period) for $0\degree\leq\theta\leq30\degree$. Since TMDC monolayers belong to $D_{3h}$ point group, the moiré superlattice constant can be calculated using the formula: $a_{moir\Acute{e}}=0.5a/sin(\theta/2)$, when $0\degree\leq\theta\leq30\degree$ and $a_{moir\Acute{e}}=0.5a/sin(30\degree-\theta/2)$, when $30\degree\leq\theta\leq60\degree$  \cite{moon2013optical}, where $a$ is the in-plane lattice constant and $\theta$ is the relative twisting angle between layers. As shown in Fig. \ref{fig:1_tB_Raman}(c), $I_{B_{2g}}/I_{{E_{2g}}/{A_{1g}}}$ follows the progression of the moiré period with twisting angle very nicely. To further confirm this correlation, Raman spectra of such tB systems are also simulated by theoretical calculations. Unlike the intensity, there are no significant changes in the frequencies of the $B_{2g}$, $A_{1g}$ and $E_{2g}$ modes observable with twisting angle in both experimental and theoretical results. The phonon modes in the twisted structures can be viewed as off-center phonons in the monolayer \cite{Lin2018}. Therefore, the process of twisting is similar to mapping the phonon dispersion of the monolayer WSe$_2$ in the in-plane direction. All three phonon modes considered here are high-frequency ones, \textit{i.e.} mainly strong and localized valence bonds are determinant for the frequencies. These phonon modes thus reveal weak dispersion in the in-plane direction and consequently the resulting phonon frequencies hardly change upon twisting.

In Fig. \ref{fig:1_tB_Raman}(c), the simulated intensity ratio of $I_{B_{2g}}/I_{{E_{2g}}/{A_{1g}}}$ shows a similar trend as the experimental one. This trend further confirms that the periodic change of the intensity ratio is  closely related to the geometric arrangement of the atoms. However, compared to the experimental result, the intensity ratio is markedly weaker, which is likely due to the inability of taking the resonant Raman conditions into account theoretically. It is noteworthy that the $B_{2g}$ mode can be observed under resonant Raman conditions even in some configurations, in which it is forbidden by symmetry selection rules \cite{luo2013effects}. The complete simulation results of the Raman spectra as a function of different twisting angles can be found in the supporting information.

To explain the interesting behavior of the $I_{B_{2g}}/I_{{E_{2g}}/{A_{1g}}}$ intensity ratio, we propose the hypothesis that the change of intensity ratio is induced by the highly localized stacking structure, which results in different interlayer coupling. A schematic diagram of moiré superlattices with different twisting angles is shown in Fig. S5. One can clearly identify that with increasing twisting angle the area of the energetically favorable AB stacking decreases in a moiré superlattice as also confirmed in Ref. \cite{quan2021phonon}. As seen in Fig. S5 (d), no more AB stacking exists in such superlattice when $\theta$ = 30°. We thus predict that the lack of AB stacking causes reduced interlayer coupling and the two layers cannot maintain 180° vibrational phase difference. In other words, in a 30° tB WSe$_2$ the two contributing monolayers vibrate like two independent monolayers.

\begin{figure} 
\makebox[\textwidth][c]{\includegraphics[width=1\textwidth]{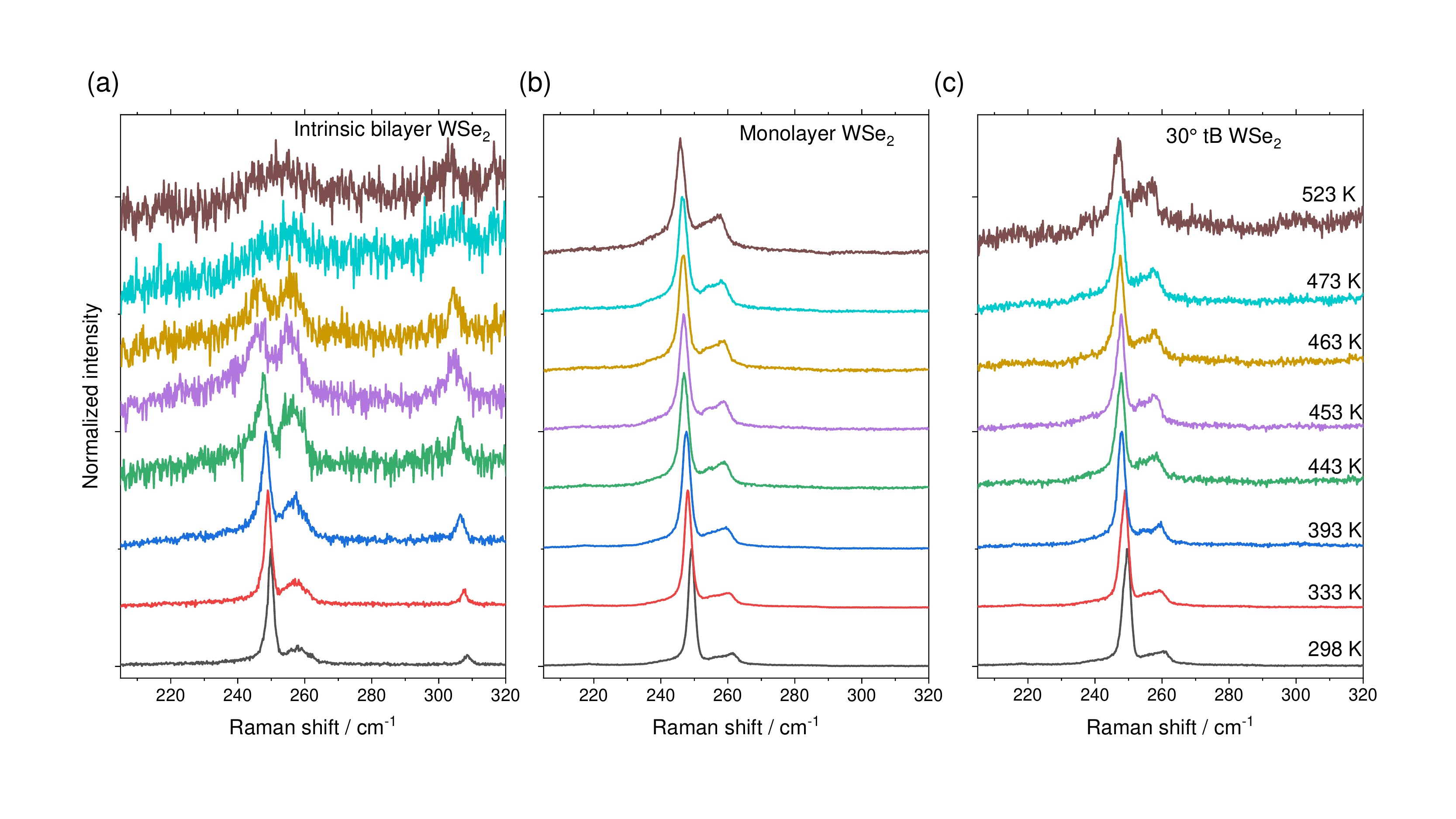}}  
\caption{Temperature-dependent Raman spectra of (a) iB, (b) monolayer, (c) 30° tB WSe$_2$.}  
\label{fig:2_Temperature_Raman}
\end{figure}

\begin{figure} [!htb]
\makebox[\textwidth][c]{\includegraphics[width=1\textwidth]{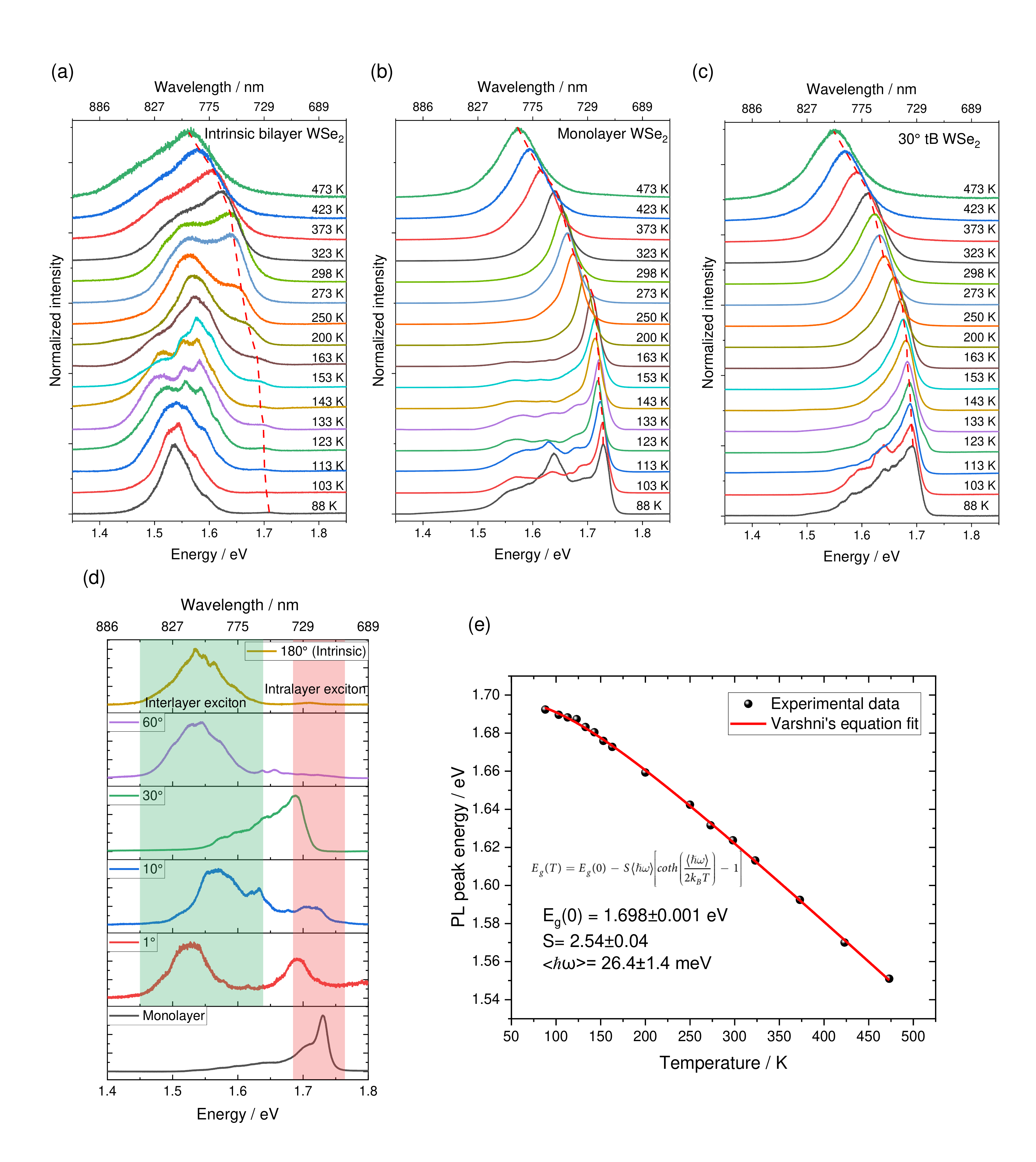}}  \caption{Temperature-dependent PL spectra of (a) iB (b) monolayer and (c) 30° tB WSe$_2$. (d) PL spectra of monolayer, 1°, 10°, 30°, 60° tB, and iB WSe$_2$ at 88 $K$. (e) A exciton (PL) energy from 30° tB WSe$_2$ as a function of temperature and fitting with Varshni’s equation.}  
\label{fig:3_Temperature_PL}
\end{figure}

To further confirm this hypothesis, we performed a series of temperature-dependent Raman measurements because the optical response of WSe$_2$ is highly influenced by the excitonic resonance condition \cite{del2014excited}. Temperature-dependent Raman spectra for iB, monolayer, and 30° tB WSe$_2$ are shown in Fig. \ref{fig:2_Temperature_Raman} (for simplicity, selected Raman spectra are shown in the main text, more Raman spectra can be found in Fig. S6). With increasing temperature, Raman features for iB, monolayer and 30° tB WSe$_2$ shift towards lower frequency together with a broadening of the FWHM,  which is in good agreement with what was reported in Refs. \cite{dos2020temperature,li2020temperature}. The most interesting observation in Fig. \ref{fig:2_Temperature_Raman}(a) is that the Raman intensity of iB WSe$_2$ decreases drastically with increasing temperature. The signal to noise ratio becomes so low that the Raman features are almost invisible when the temperature is higher than 473 K, whereas for the monolayer and 30° tB WSe$_2$ in Fig. \ref{fig:2_Temperature_Raman}(b, c) the Raman spectra remain intense at high temperature. The $E_{2g}$, $A_{1g}$ vibrational modes in WSe$_2$ are sensitive to the resonance condition with the $A’$ exciton, where the $A’$ arises from the splitting of the ground and excited states of the $A$ transition (K-K transition in the Brillouin zone) due to the intralayer perturbation of the $d$ electron band by the Se $p$ orbitals \cite{zhao2013evolution,beal1972transmission,bromley1972band}. However, information about the $A’$ exciton energy in tB WSe$_2$ systems is still lacking in the literature. We therefore run a temperature-dependent PL measurement to determine the $A$ exciton energy, which is used to derive the $A’$ exciton energy. It is noteworthy that the Stokes shift for WSe$_2$ is $3-5 meV$ \cite{zhao2013evolution,tran2017disorder}, which is negligible for the following calculation, thus the PL peak position is taken as the exciton transition energy. Temperature-dependent PL spectra of iB, monolayer, and 30° tB WSe$_2$ are shown in Fig. \ref{fig:3_Temperature_PL}(a-c). The peaks guided by red dashed line correspond to the $A$ exciton PL emission. The peak positions of the $A$ exciton in 30° tB WSe$_2$ are plotted in Fig. \ref{fig:3_Temperature_PL}(e) and are fitted to a modified Varshni’s equation (for iB and monolayer see Fig. S7 (d) and (e)), which describes the temperature dependence of a semiconductor bandgap \cite{huang2016probing,ross2013electrical,tongay2012thermally,o1991temperature}:

\begin{equation}
    E_g(T)=E_g(0)-S\left \langle \hbar\omega \right \rangle	\left[coth\left(\frac{\left \langle \hbar\omega \right \rangle}{2k_BT}\right)-1 \right]
\end{equation}

where $E_g(0)$ is the transition energy at $T=0 K$, $S$ is a dimensionless constant that describes the strength of electron-phonon coupling and $\left \langle \hbar\omega \right \rangle$ describes the average acoustic phonon energy involved in electron-phonon interactions.

For iB WSe$_2$, the $A$ exciton energy $E_{A,iB} (298~K)=1.64~eV$, $E_{A,iB} (473 ~K)=1.56~eV$, and the energy difference between 473 K and 298 K $\Delta E_{A,iB}=0.08~eV$, the $A’$ exciton energy $E_{A',iB} (298~K)=2.28~eV$\cite{zhao2013evolution,del2014excited}, $E_{A',iB} (473~K)=E_{A',iB} (298~K)-\Delta E_{A,iB}=2.20~eV$. Since strong resonance enhancement only occurs in an excitation energy range approximately $\pm 0.1~eV$ from the $A’$ exciton energy \cite{del2014excited}, our excitation energy of $E_{laser}=2.41~eV$ is far from the resonance condition at 473 K and results in strongly decreasing intensity in the Raman spectra of Fig. \ref{fig:2_Temperature_Raman}(a). Following the same procedure for monolayer WSe$_2$, $E_{A',1L} (473~K)=2.34~eV$, which still fulfills the resonance condition. Therefore, the Raman spectra remain intense at 473 K in Fig. \ref{fig:2_Temperature_Raman}(b). For 30° tB WSe$_2$, $E_{A,30^{\circ} tB} (298~K)=1.62~eV$, $E_{A,30^{\circ} tB} (473~K)=1.55~eV$, and $\Delta E_{A,30^{\circ} tB}=0.07~eV$. However, the information about the $A’$ exciton energy is still lacking. Therefore, we take the values of the energy difference between the $A’$ and $A$ excitons in both iB and monolayer WSe$_2$ for the calculation. When taking $\Delta E_{A'-A,30^{\circ} tB}=\Delta E_{A'-A,iB}=0.70~eV$, following the previous procedure, $E_{A',30^{\circ} tB} (473~K)=2.25~eV$, which is far away from the resonance condition and disagrees with our experimental observation. However, when taking $\Delta E_{A'-A,30^{\circ} tB}=\Delta E_{A'-A,1L}=0.74~eV$, which results in $E_{A',30^{\circ} tB} (473~K)=2.29~eV$, so that the excitation energy is a little further away from the resonance condition than in monolayer, but is still close enough to cause resonance enhancement. This is in good agreement with our experimental results in Fig. \ref{fig:2_Temperature_Raman}(c), namely that we can still observe the Raman features from 30° tB WSe$_2$ at 473 K well, but the spectra have relative lower signal-to-noise ratio compared to the monolayer results in Fig. \ref{fig:2_Temperature_Raman} (b). It is therefore evident from the resonant Raman scattering point of view that the Raman spectra of 30° tB WSe$_2$ resemble those of the monolayer indicating that the two layers behave like 2 independent monolayers. Moreover, Fig. \ref{fig:3_Temperature_PL}(d) displays the PL spectra of tB WSe$_2$ with different twisting angle. One can clearly observe the interlayer transition in iB and 1°, 10°, 60° tB WSe$_2$. Yet the interlayer transition seems to be absent in 30° tB WSe$_2$, which also agrees with our hypothesis that 30° tB WSe$_2$ has a similar optical response as 2 independent monolayers.

\subsection{Twisted hetero-bilayers}

Furthermore, we carry out similar Raman investigations on twisted hetero-bilayer (thB) TDMCs. Inspired by the tear-and-stack method, we developed an advanced tear-and-stack method, which enables to prepare samples with different crystal orientation (details described in Fig. S2) on the same substrate. This method offers the possibility to carry out large area spectroscopic mapping and to spatially resolve and study the optical properties of 2D materials.

\begin{figure}[!htb]
\makebox[\textwidth][c]{\includegraphics[width=1\textwidth]{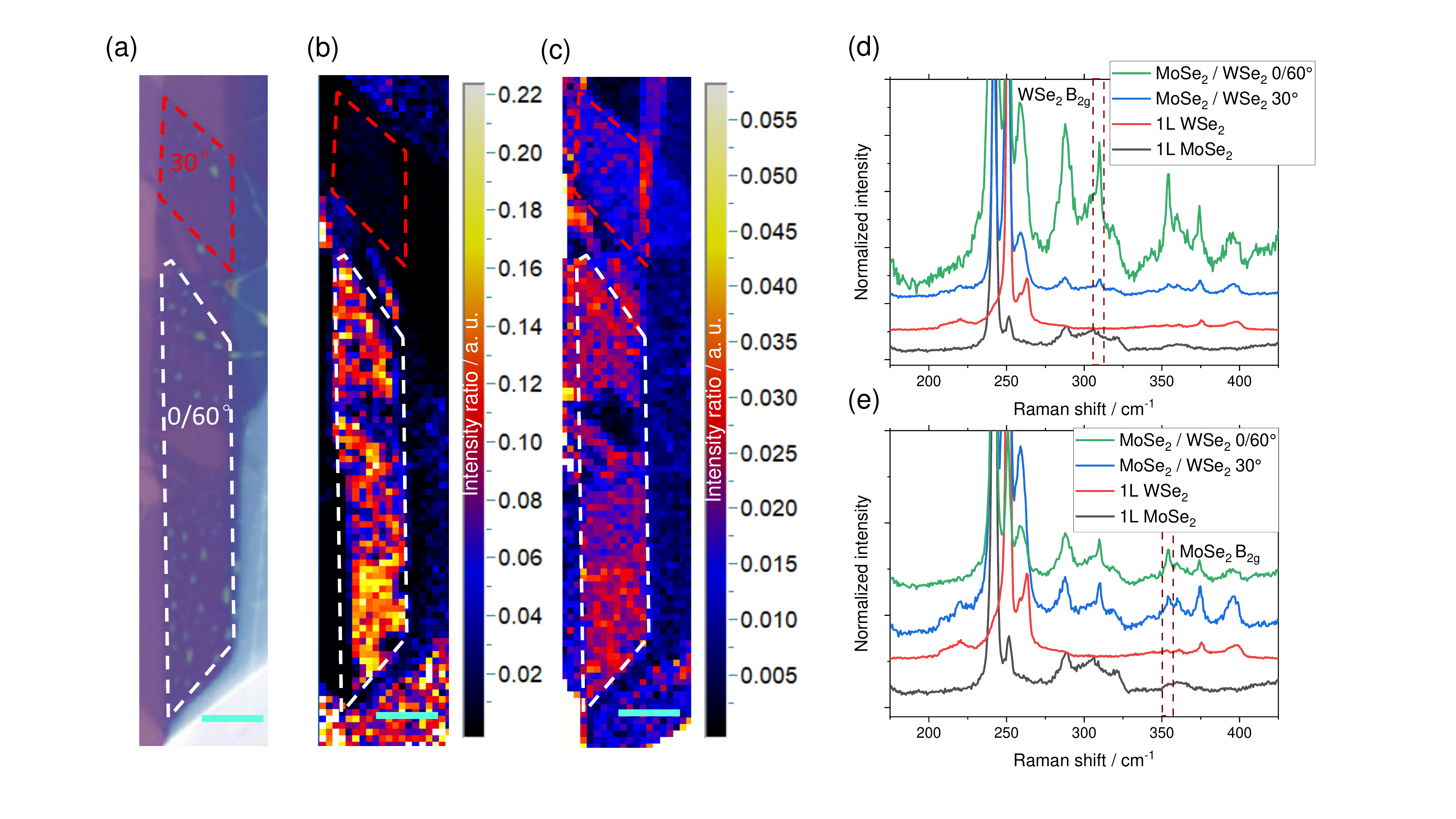}}  \caption{Raman map of twisted MoSe$_2$/WSe$_2$ hetero-bilayer. (a) optical image of twisted MoSe$_2$/WSe$_2$, areas marked by red and white dashed lines have twisting angles of 30° and 0/60°, respectively. Raman intensity ratio map of (b) WSe$_2$ $I_{B_{2g}}/I_{{E_{2g}}/{A_{1g}}}$ and (c) MoSe$_2$ $I_{B_{2g}}/I_{A_{1g}}$. Corresponding Raman spectra normalized to (d) WSe$_2$ $E_{2g}$/$A_{1g}$ and (e) MoSe$_2$ $A_{1g}$. Scalebar is 5 \textmu m.}  
\label{fig:4_Raman_map}
\end{figure}

The optical image of thB MoSe$_2$/WSe$_2$ prepared by such method is shown in Fig. \ref{fig:4_Raman_map}(a). The areas marked by red and white dashed lines have twisting angles of 30° and 0/60°, respectively. Figs. \ref{fig:4_Raman_map}(b) and (d) display the Raman intensity ratio map of WSe$_2$ $I_{B_{2g}}/I_{{E_{2g}}/{A_{1g}}}$ and corresponding Raman spectra normalized with respect to the WSe$_2$ ${E_{2g}}/{A_{1g}}$ peak. The WSe$_2$ $B_{2g}$ peak is marked by a dashed line in Fig. \ref{fig:4_Raman_map}(d), which is absent in monolayer and relatively weak in the 30° thB MoSe$_2$/WSe$_2$, while becomes very intense in 0/60° thB. The strong contrast can be visualized in the Raman intensity map in Fig. \ref{fig:4_Raman_map}(b). The argument still holds in Fig. \ref{fig:4_Raman_map}(c) when the Raman intensity ratio map of MoSe$_2$ $I_{B_{2g}}/I_{A_{1g}}$ is plotted and in Fig. \ref{fig:4_Raman_map}(e) when the spectra are normalized with respect to the MoSe$_2$ $A_{1g}$ peak. Similar experiments were also performed on different combinations of materials such as WSe$_2$/MoS$_2$ and MoSe$_2$/MoS$_2$. As shown in Fig. S8, all experiments provide similar results. Our study thus provides quite universal results for all kinds of Mo- and W-based TMDC thB.

\section{Conclusions}

In summary, we investigated the high-frequency Raman modes of tB WSe$_2$ both experimentally and theoretically. We observed that the intensity ratio $I_{B_{2g}}/I_{{E_{2g}}/{A_{1g}}}$  follows the same periodic trend as the moiré superlattice constant and can be used as a signature of the lattice dynamics in twisted 2D systems. Focusing on 30° tB WSe$_2$, where $I_{B_{2g}}/I_{{E_{2g}}/{A_{1g}}}$  is almost zero, temperature-dependent Raman and PL measurements were performed. 30° tB WSe$_2$ is proven to have a similar optical response as two independent monolayers indicating that the interlayer coupling is extremely weak. This further confirms our hypothesis that the high-frequency Raman intensity ratio $I_{B_{2g}}/I_{{E_{2g}}/{A_{1g}}}$ can be considered as the signature of the lattice dynamics in tB WSe$_2$ and the highly localized interlayer coupling. Finally, we explored different combination of Mo- and W-based TMDC thB systems to establish a universal behavior of interlayer coupling in these systems, which is fundamental for moiré physics and device applications.

\section{Acknowledgments}
The authors gratefully acknowledge  financial support by the Deutsche Forschungsgemeinschaft (DFG) for projects ZA 146/43-1, ZA 146/44-1, and ZA 146/47-1.

\section{Author contributions}
Y.P., M.R. and I.M. performed the optical measurements. Y.P. fabricated the samples. S.L. conducted the theoretical calculation and Raman simulation. M.R. and D.R.T.Z. supervised the work and involved in the evaluation and interpretation of the results. Y.P. and S.L. wrote the manuscript. All authors contributed to the understanding of results and data analysis.

\section{Conflict of interest}
The authors declare no conflict of interest.

\printbibliography


\section*{Supplementary material}

# Supplementary information: Signature of lattice dynamics in twisted 2D homo/hetero-bilayers

Yang Pan[1,2], Shutong Li[3], Mahfujur Rahaman[1,2]*, Ilya Milekhin[1,2], and Dietrich R. T. Zahn[1,2]*

[1]Semiconductor physics, Institut für Physik, Chemnitz University of Technology, Chemnitz, Germany

[2]Center for Materials, Architectures, and Integration of Nanomembranes (MAIN), Chemnitz University of Technology, Chemnitz, Germany

[3]Department of Chemical Engineering and Materials Science, University of Minnesota, Minneapolis, Minnesota, USA

*Corresponding author: Dietrich R. T. Zahn: zahn@physik.tu-chemnitz.de

Mahfujur Rahaman

## Sample preparation

### Preparation of twisted homo-bilayers:

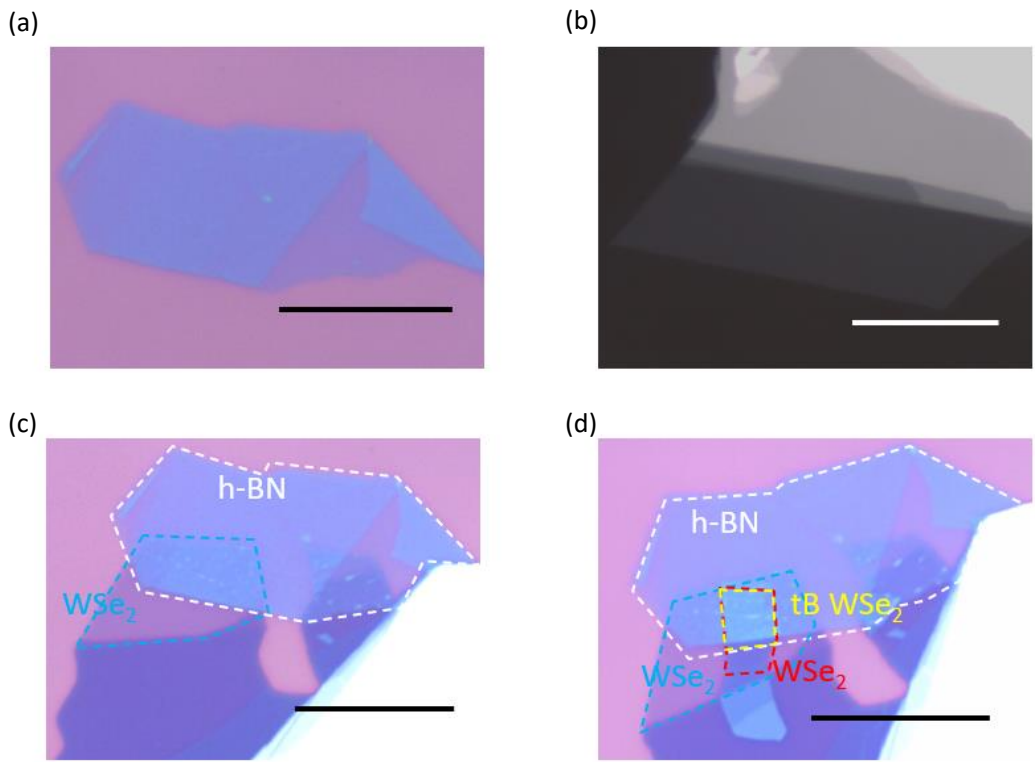

Figure S1. tB WSe$_2$ sample preparation. Optical image of (a) few layer hBN exfoliated on SiO$_2$ substrate, (b) monolayer WSe$_2$ exfoliated on PDMS, (c) partially transferred monolayer WSe$_2$ on hBN, and (d) tB WSe$_2$ on hBN. Scale bar is 20 μm.

## Preparation of twisted hetero-bilayers

30° tB WSe$_2$ is prepared following a tear-and-stack method. Due to the crystal symmetry, the breaking edges of these 2D materials can only be armchair or zigzag (shown in Fig. S2 (d)). A MoSe$_2$ monolayer is later transferred on the 30° tB WSe$_2$, with one edge aligned to any WSe$_2$ edge at an angle of $30° \times n$ (n is any integer number). This alignment results in 2 areas of heterostructures, one with a relative twisting angle of 30° and the other with 0° or 60°.

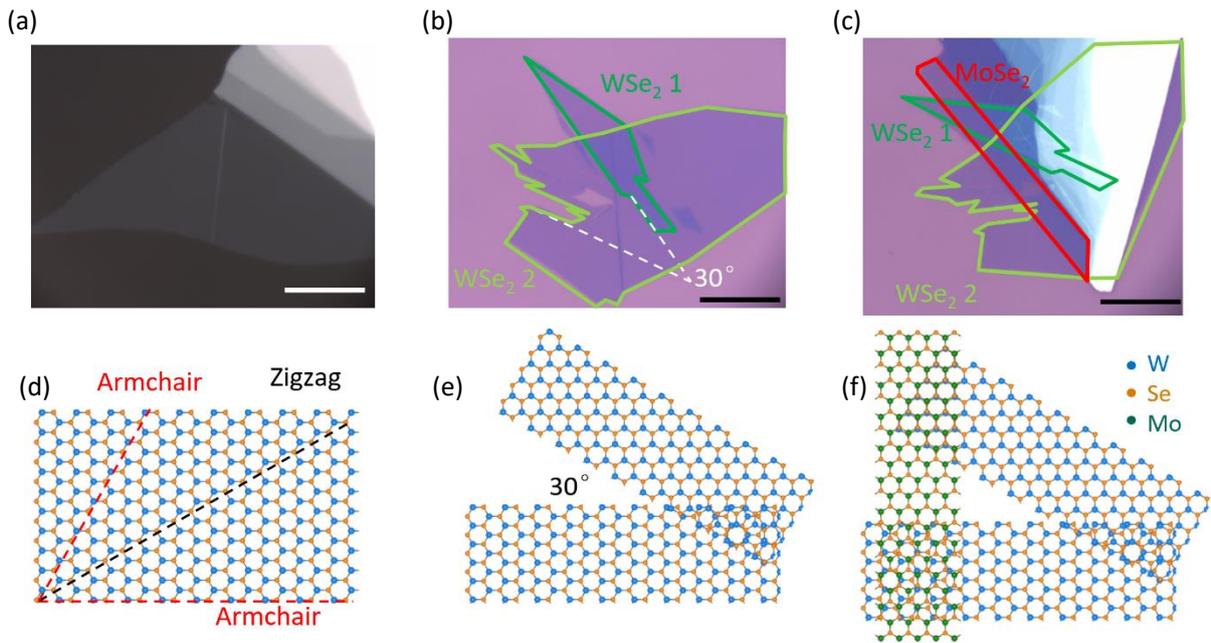

Figure S2. Twisted hetero-bilayer sample preparation. Optical image of (a) monolayer WSe$_2$ exfoliated on PDMS, (b) 30° tB WSe$_2$ on SiO$_2$, (c) twisted MoSe$_2$/WSe$_2$ hetero-bilayer. (d)-(e) are the schematic crystal structures, respectively. Scalebar is 10 μm.

# Atomic structures for first-principles calculations

Bulk WSe$_2$ is known to crystallize in the 2H structure. However, the artificial stacking of 0° and 60° twisted bilayer WSe$_2$ can lead to at least 5 possible high symmetry atomic structures, as shown in Fig. S3.

First-principles calculations are performed for these high symmetry structures: the 2H, 3R phases at 0° and AB', A'B, 2H-phases at 60° to find the ground state structures.

A set of commensurate twisted bilayer WSe$_2$ structures are also calculated with various numbers of atoms, including those with twisting angles of 21.8°, 27.8° and 38.2°. These degrees are specifically selected in order to apply the periodic boundary condition.

It is found that the 2H phase at 60° and the 3R phase at 0° have the lowest energy as shown in Fig. S3. Thus, these two phases will be selected as unit cells for the simulations of 0° and 60° twisted bilayers. Structures at 21.8°, 27.8° and 38.2° are also fully relaxed for the further simulations, using density functional theory. We pick these three data points because they are representative and close to 30°, although the demand for computational resources is very significant.

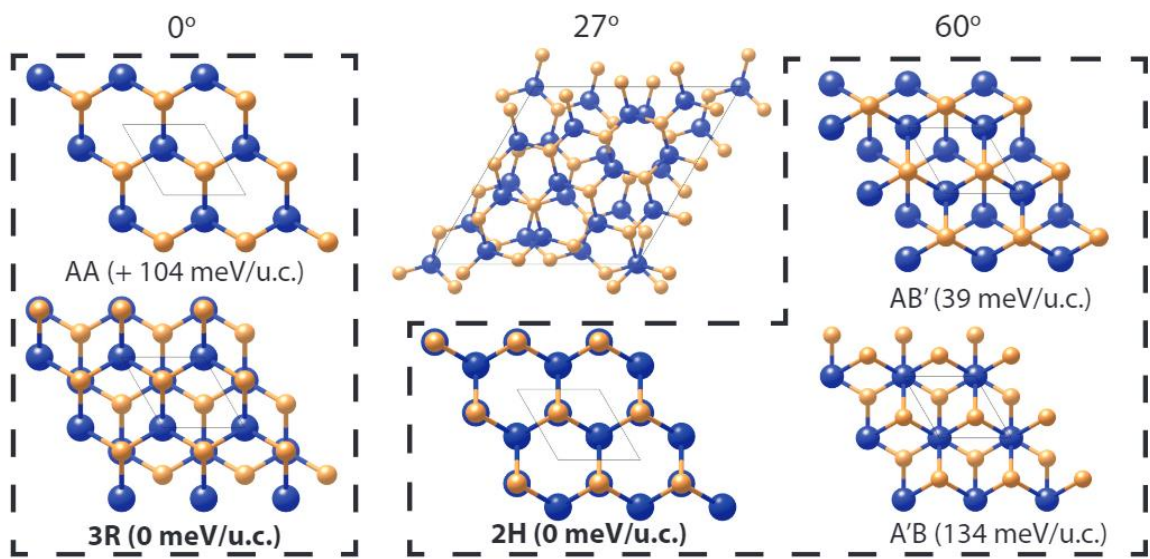

Figure S3. Possible structures of twisted bilayer WSe$_2$. The 3R phase has the lowest energy when the bilayer is untwisted. While at 60°, the 2H phase is the most stable one. Other possible crystal structures for 0° and 60° twisting angle are also shown in this plot, they are possibly observable when mechanically stacked. The commensurate crystal structures used for 27° twisting angle are also shown in the middle. The same calculations were also performed on 21° and 38° twisted structures.

# Raman spectra simulation

The simulation configurations are exactly the same as the ones used for structure relaxation. The calculated frequencies are listed in the main text Table 1. The phonon modes are extracted and later used for the Raman tensor calculation. The off-resonant Raman tensor can be calculated by the change of polarizability:

$$\tilde{R}_{\alpha\beta} \propto \frac{\Delta \varepsilon^{\infty}_{\alpha\beta}}{\Delta Q(s)} = \frac{1}{2}\left(-\frac{\varepsilon^{\infty}_{\alpha\beta}(-s)}{\Delta Q(s)} + \frac{\varepsilon^{\infty}_{\alpha\beta}(+s)}{\Delta Q(s)}\right)$$

where the $\varepsilon$ stands for the ion-clamped dielectric constant, *i.e.* the dielectric constant at infinite high frequency. The $Q(s)$ stands for the amplitude of normal mode s and $\alpha$, $\beta$ indicate the directions. The normal mode *s* comes from the phonon calculation mentioned above. The derivative of them can be calculated by freezing in the phonon mode *s* in both positive and negative directions and calculate the change of dielectric constant then divided by the amplitudes.

We use the Perturbation Expression After Discretization (PEAD) method to determine the dielectric constant at infinity frequencies (which is the ion-clamped dielectric constant). This method is a self-consistent calculation under an applied electric field. The applied electric field was set to be 10 meV/Å.

Later on, the non-resonant Raman intensity can be determined by the Raman tensor: $I_{Raman} \propto |\boldsymbol{e_i} \cdot \tilde{R} \cdot \boldsymbol{e_s}|$. The $\boldsymbol{e_i}$ and $\boldsymbol{e_s}$ are the polarization direction of incident and scattered light. In our experimental setup, the incident light is polarized to the in-plane direction. Both directions of in-plane scattered light were collected, which results in the total Raman intensity $I \propto I_{xx} + I_{yy} \propto R_{xx}^2 + R_{yy}^2$. Fig. S4 shows the calculated Raman spectra when the bilayer $WSe_2$ is twisted from 0° to 60°. The spectra in rectangular box are zoomed in by 100 times for visualization. However, compared to the experimental results, the $B_{2g}$ mode is weaker, which is likely due to the inability of capturing the resonant Raman intensity. It is reported that the $B_{2g}$ mode has resonant-Raman activity and can be observed even in some selection-rule prohibited structures[1].

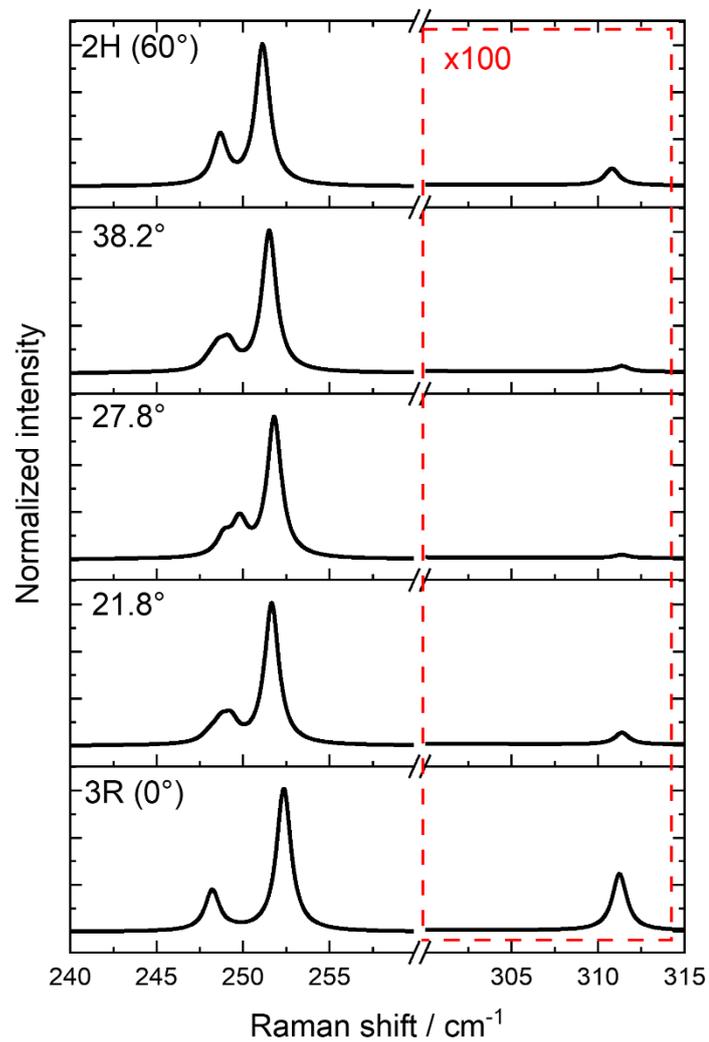

Figure S4. Simulated Raman spectra of 3R (0°), 21.8°, 27.8°, 38.2° and 2H (60°) bilayer $WSe_2$. The spectra between 300 $cm^{-1}$ to 315 $cm^{-1}$ in the rectangular box are zoomed by 100 times for better visualization.

# Crystal structure in moiré superlattices

The structures of 2°, 5°, 10°, and 30° tB WSe$_2$ are shown in Fig. S5. The moiré superlattice constant is calculated with the equation[2]:

$$a_{moiré} = \frac{0.5a}{sin(\theta/2)}$$

Where $a$ is the in-plane lattice constant of WSe$_2$ and $\theta$ is the twisting angle. The energy preferable AB and BA stacking area is absent in 30° tB WSe$_2$.

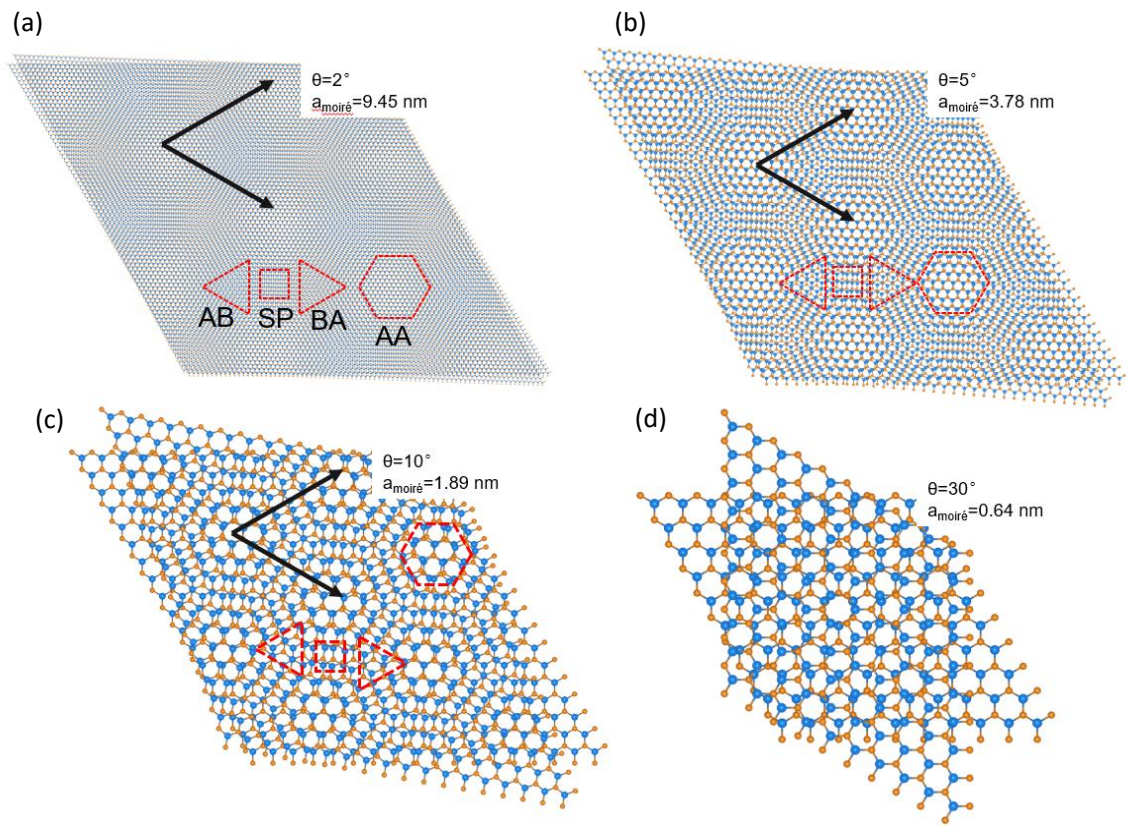

Figure S5. Schematic diagrams of WSe$_2$ moiré superlattices with different twisting angles: (a) 2°, (b) 5°, (c) 10°, and (d) 30°.

## Temperature-dependent Raman spectra of WSe$_2$

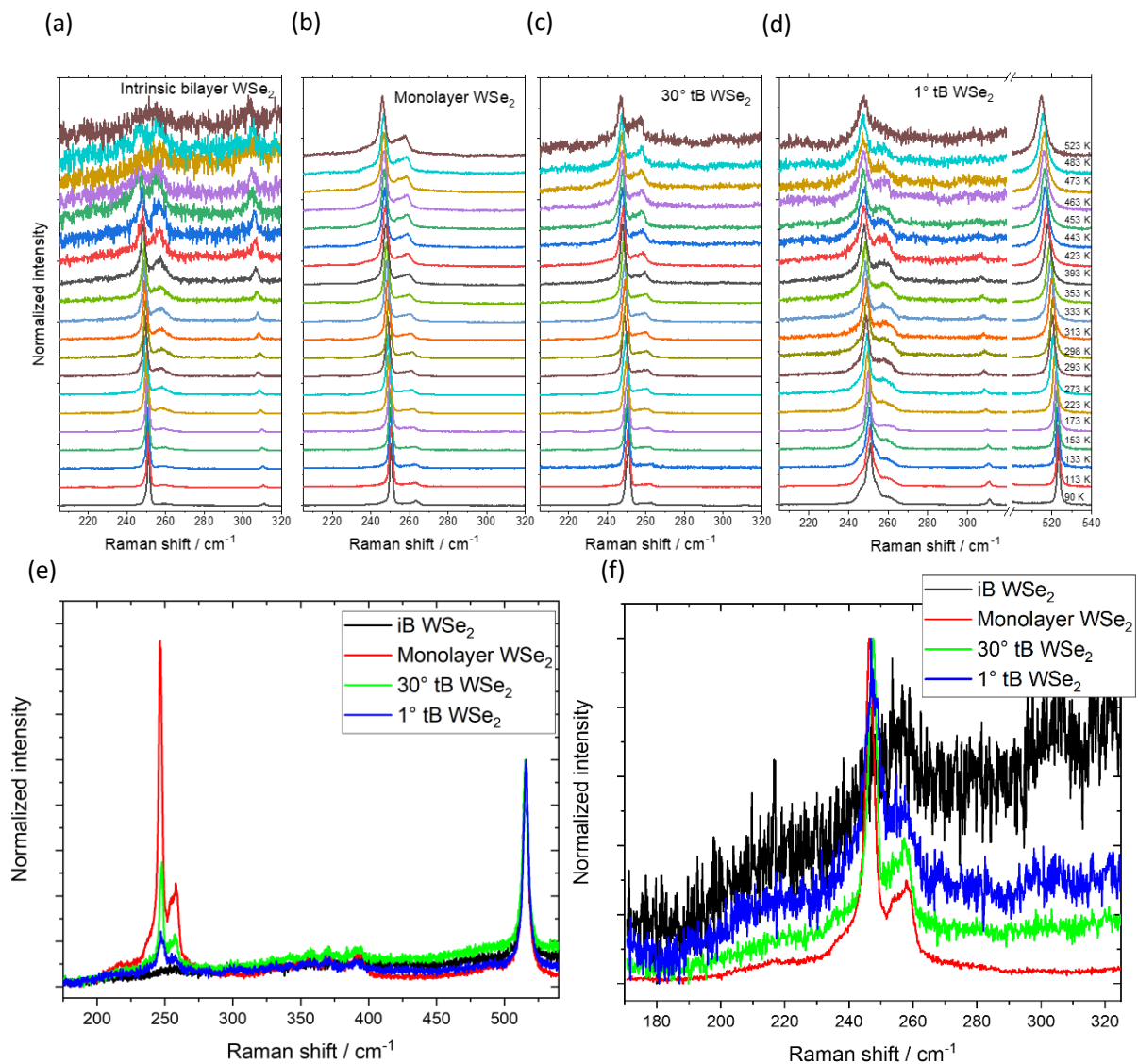

Figure S6. Temperature-dependent Raman spectra of (a) iB, (b) monolayer, (c) 30° tB WSe$_2$, and (d) 1° tB WSe$_2$ with the silicon Raman peak used for temperature determination. Note that the Raman spectra in the ranges of 205- 320 cm$^{-1}$ and 500- 540 cm$^{-1}$ are normalized separately. (e) and (f) are the Raman spectra of iB, monolayer, 30° tB, and 1° tB WSe$_2$ at 473 K normalized with respect to the Si and WSe$_2$ E$_{2g}$/A$_{1g}$ peaks, respectively.

Temperature-dependent Raman spectra of iB, monolayer, 30° and 1° tB WSe$_2$ are shown in Fig. S6 (a-d), respectively. The Raman signal of iB WSe$_2$ is almost not visible when the temperature is higher than 473 K, while monolayer and 30° tB WSe$_2$ behave completely differently from the iB case, whereas 30° tB WSe$_2$ has similar behavior as monolayer. The Raman signal of 1° tB WSe$_2$ is still visible at high temperature. However, as shown in Fig. S6 (e) and (f), it has much lower intensity and signal-to-noise ratio compared to monolayer and 30° tB WSe$_2$, which can be explained by relatively weaker interlayer coupling because of the existence of non-AA/AB stacking area and hydrocarbonate contamination at the interface due to the artificial stacking.

The temperature dependent Raman spectra of Si are also plotted in Fig. S6(d). With increasing temperature, the Si Raman feature shifts towards lower frequency and shows a broadening of the FWHM, which is in good agreement with previous studies[3,4]. The temperature dependence of the Si Raman peak can be used as an indicator of the temperature.

## Temperature-dependent PL spectra and Varshni's plot

Temperature-dependent PL spectra of iB, monolayer, and 30° tB WSe$_2$ are shown in Fig. S7 (a-c). The 30° tB WSe$_2$ has a similar PL response as the monolayer WSe$_2$, while there is a huge difference to that of iB WSe$_2$. The direct transition PL peak position is plotted in Fig (d-e) and fitted with the modified Varshni's equation[5–8]:

$$E_g(T) = E_g(0) - S\langle\hbar\omega\rangle\left[\coth\left(\frac{\langle\hbar\omega\rangle}{2k_BT}\right) - 1\right]$$

Where $E_g(0)$ is the transition energy at T=0 K, $S$ is a dimensionless constant that describes the strength of electron-phonon coupling, and $\langle\hbar\omega\rangle$ describes the average acoustic phonon energy involved in electron-phonon interactions.

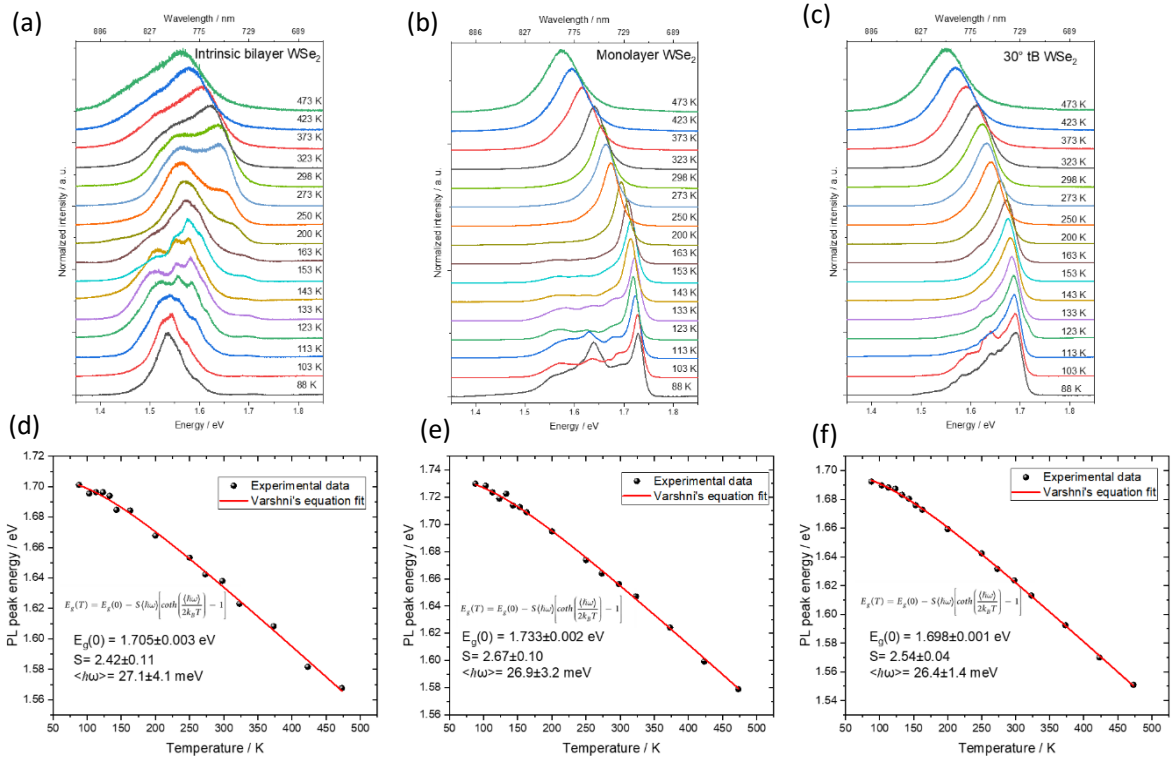

Figure S7. Temperature dependent Raman spectra of (a) iB, (b) monolayer, (c) 30° tB WSe$_2$ and the corresponding A exciton positions fitted with Varshni's equation (d)-(f).

# Raman map of twisted hetero-bilayers

We further investigated different combination of TDMCs. Both WSe$_2$/MoS$_2$ and MoSe$_2$/WSe$_2$ hetero-bilayers show consistent results as in main text. One can clearly identify the contrast between the two areas with different crystal orientations in the Raman intensity ratio map.

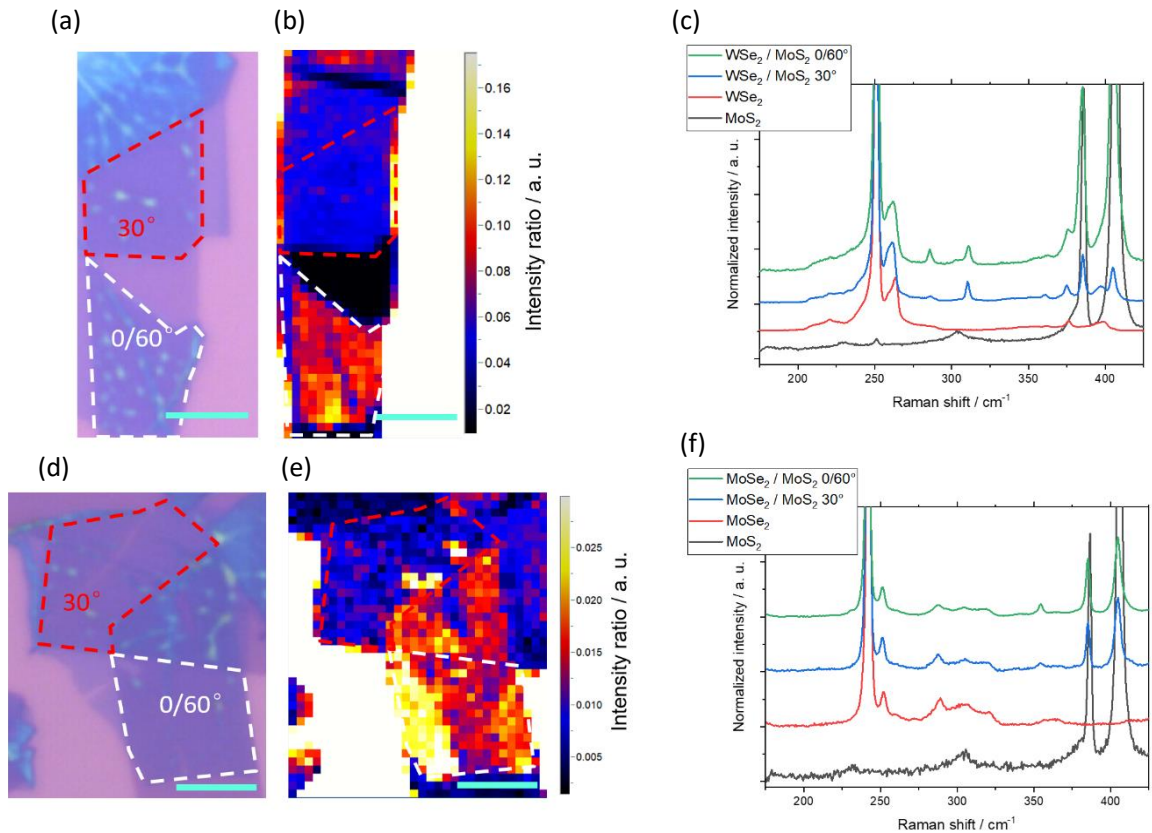

Figure S8. Raman maps of twisted hetero-bilayers. Optical image of twisted (a) WSe$_2$/MoS$_2$ (d) MoSe$_2$/WSe$_2$, areas marked by red and white dashed lines have twisting angles of 30° and 0/60°, respectively. Raman intensity ratio maps of (b) WSe$_2$ $I_{B_{2g}}/I_{E_{2g}\ A_{1g}}$ and (e) MoSe$_2$ $I_{B_{2g}}/I_{A_{1g}}$. Corresponding Raman spectra normalized to (c) WSe$_2$ E$_{2g}$/A$_{1g}$ and (f) MoSe$_2$ A$_{1g}$. Scalebar is 5 µm.


\end{document}